%% file: conference_101719.tex
\documentclass[conference]{IEEEtran}
\IEEEoverridecommandlockouts
\usepackage{cite}
\usepackage{amsmath,amssymb,amsfonts}
\usepackage{algorithmic}
\usepackage{textcomp}
\usepackage{xcolor}
\usepackage{multirow}
\usepackage{comment}
\usepackage{booktabs, soul}
\usepackage{tabularx, color, soul}
\usepackage[hidelinks]{hyperref}
\usepackage{pifont}% http://ctan.org/pkg/pifont
\usepackage[acronym,shortcuts,acronymlists={hidden}]{glossaries}
\input{glossary.tex}
\def\BibTeX{{\rm B\kern-.05em{\sc i\kern-.025em b}\kern-.08em
    T\kern-.1667em\lower.7ex\hbox{E}\kern-.125emX}}
    
\newcolumntype{C}{>{\centering\arraybackslash}X}
\usepackage{caption,subcaption,graphicx}
\usepackage{pifont}% http://ctan.org/pkg/pifont
\begin{document}

\title{CAESR: Conditional Autoencoder and Super-Resolution for Learned Spatial Scalability
\vspace{-0.15in}}
\pdfminorversion=5
\pdfobjcompresslevel=2
\makeatletter
\newcommand{\linebreakand}{%
  \end{@IEEEauthorhalign}
  \hfill\mbox{}\par
  \mbox{}\hfill\begin{@IEEEauthorhalign}
}
\makeatother

\author{\IEEEauthorblockN{Charles Bonnineau$^{\star \dagger \ddagger}$, Wassim Hamidouche$^{\star \ddagger}$, Jean-Fran\c cois Travers$^\dagger$, Naty Sidaty$^\dagger$, \\ Jean-Yves Aubié$^\star$ and Olivier Deforges$^\ddagger$}
\IEEEauthorblockA{$^\star$IRT b$<>$com, Cesson-Sevigne, France, \\
$^\dagger$TDF, Cesson-Sevigne, France, \\
$^\ddagger$Univ Rennes, INSA Rennes, CNRS, IETR - UMR 6164, Rennes, France
\vspace{-0.15in}}}
\maketitle

%%%%%%%%% ABSTRACT
\begin{abstract}

%The enhancement-layer consists of an encoder-decoder pair trained jointly with a hyperprior and a deep \gls{SR} module. 

In this paper, we present CAESR, an hybrid learning-based coding approach for spatial scalability based on the \gls{VVC} standard. Our framework considers a low-resolution signal encoded with \gls{VVC} intra-mode as a \gls{BL}, and a deep conditional \gls{AE-HP} as an \gls{EL} model. The \gls{EL} encoder takes as inputs both the upscaled \gls{BL} reconstruction and the original image. Our approach relies on conditional coding that learns the optimal mixture of the source and the upscaled \gls{BL} image, enabling better performance than residual coding. On the decoder side, a \gls{SR} module is used to recover high-resolution details and invert the conditional coding process. Experimental results have shown that our solution is competitive with the \gls{VVC} full-resolution intra coding while being scalable. 

\end{abstract}
\vspace{-0.12in}

\glsresetall
%%%%%%%%% BODY TEXT
\section{Introduction}

Over the last years, spatial scalability has been considered as a key challenge for image and video compression. Hence, dedicated video coding standards have been developed to take advantage of the existing correlations between different versions of a signal. In the case of~\gls{SHVC} \cite{boyce2015overview}, a \gls{BL} signal (low resolution) encoded with~\gls{HEVC} is used as a reference by an inter-layer processing module to encode the \gls{EL} signal (high-resolution) with the use of \gls{HLS}. More recently, \gls{LCEVC}~\cite{maurer2020overview} proposed specific tools to encode the residual information, i.e., the difference between the original video and its compressed representation. In these approaches, all tools, including scaling and transform modules, are handcrafted and separately tuned. Therefore, they may result in a suboptimal system. 

Another way to enable spatial scalability relies on spatial resolution adaptation coding framework. In this coding scheme, illustrated at the bottom of Fig.~\ref{fig:proposed}, a downscaled representation of the source signal is encoded, transmitted, and upscaled after decoding to reach the original resolution. At low-bitrate, this process may provide better coding performance than full-resolution coding~\cite{bruckstein2003down} while enabling spatial scalability with any base-layer codec. With the recent advances in deep learning, powerful pre and post-processing models have been used for spatial resolution adaptation based on existing compression standards \cite{zhang2019vistra2, ma2019perceptually, bonnineau2020versatile, bonnineau2021}. However, some high frequencies lost during the downscaling process still cannot be recovered using single post-processing modules, making performance sensitive to the content.

On the other hand, end-to-end learning models for image and video compression were proposed using deep \glspl{AE}~\cite{toderici2015variable, balle2016end, balle2018variational, cheng2020learned, lu2019dvc}. These deep models consist of a non-linear encoder-decoder pair optimized in a completely end-to-end fashion. Thus, the whole system's components are optimally tuned together regarding a given rate-distortion trade-off driven by the loss function. Hybrid layered systems have been investigated to enhance traditional codecs using an \gls{AE} as an enhancement layer model \cite{tsai2018learning, akbari2019dsslic, lee2020hybrid}. However, those solutions are based on full resolution \gls{BL} images and do not take into consideration the spatial scalability character.

\begin{figure}[t]
\centering
\includegraphics[width=\linewidth]{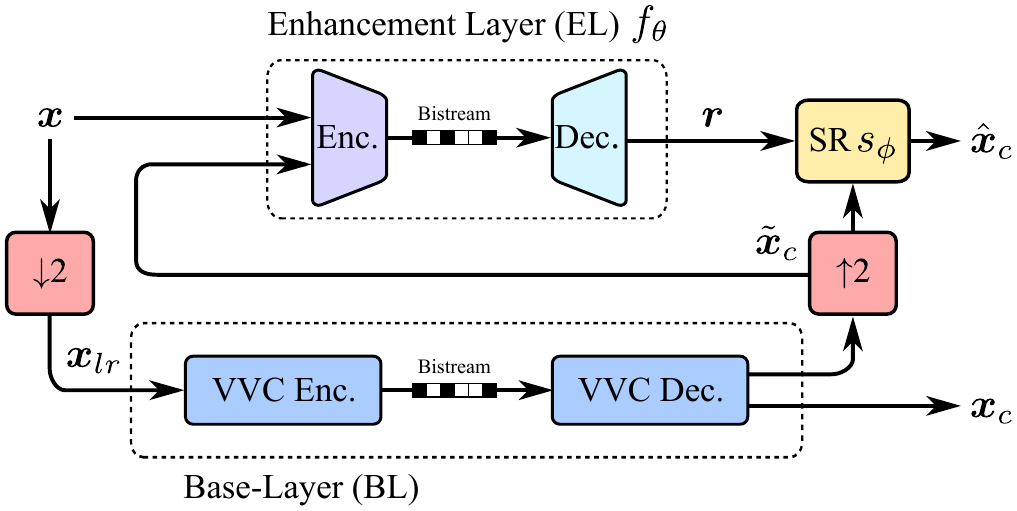}
\caption{General pipeline of CAESR. Both the downscaling and upscaling steps, denoted as $\downarrow$2 and $\uparrow$2, respectively, are performed by a handcrafted filter. On the decoder side, it allows matching both the latent residual information $\boldsymbol{r}$ and the upscaled base-layer signal $\tilde{\boldsymbol{x}}_c$ resolutions as input of the \gls{SR} module $s_\phi$.}
\label{fig:proposed}
\vspace{-0.2in}
\end{figure}

% Thus, a non-linear mixture of the source and the reconstructed \gls{BL} signal is learned, enabling better performance than residual coding.

In this paper, we present CAESR, an hybrid layered approach that uses a downscaled representation of the input image, encoded using \gls{VVC} as a \gls{BL} codec and a deep conditional autoencoder as an \glsfirst{EL} model. The key idea is to use the strong representation ability of \glspl{AE} to encode the high-resolution details lost during the downscaling and quantization steps. On the decoder side, the predicted residual information is given with the upscaled \gls{BL} signal as input to a super-resolution module to produce the reconstructed high-resolution image. We optimize the overall system by training the autoencoder jointly with the super-resolution \glsfirst{CNN} in a conditional coding scheme. To the best of our knowledge, no previous works consider spatial scalability based on the joint training of a super-resolution module and an autoencoder to transmit both coding and scaling residuals as side information. 

\begin{figure*}[t]
\begin{subfigure}[t]{\linewidth}
\centering
\includegraphics[width=\linewidth]{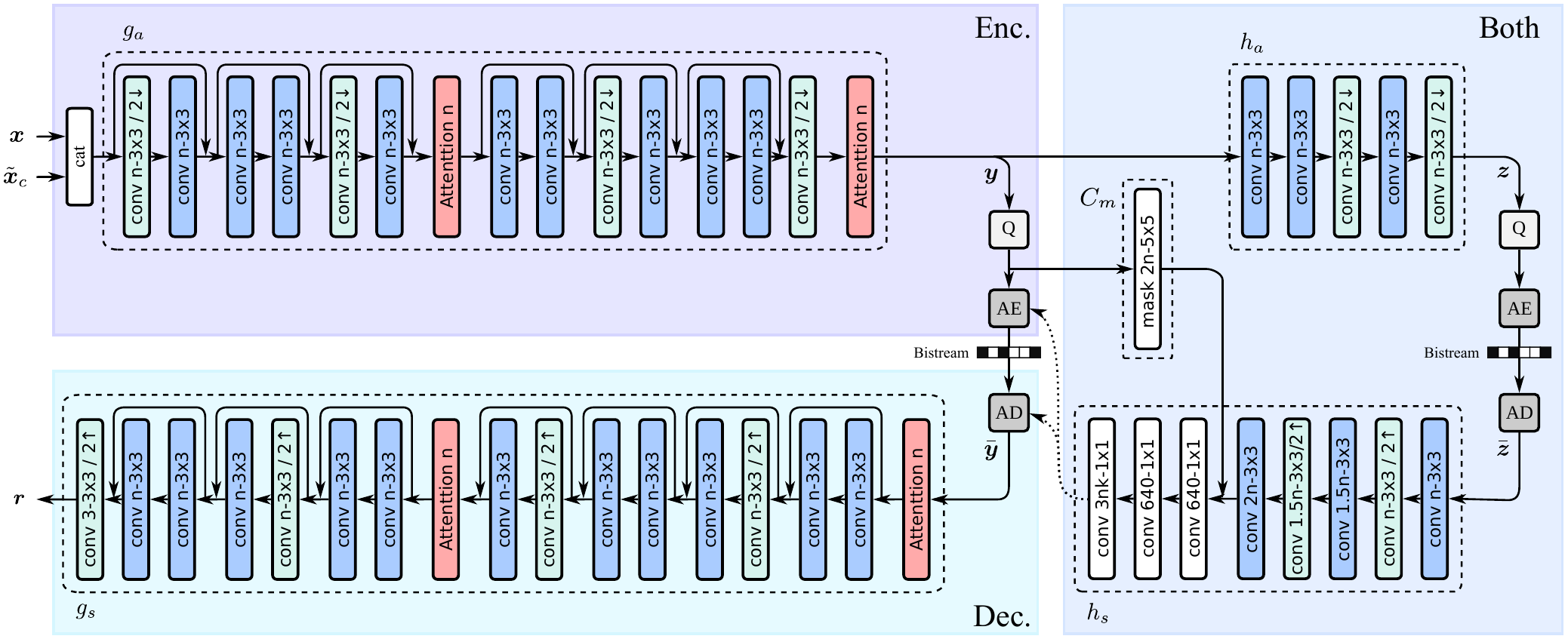}
\caption{Architecture details of the conditional autoencoder $f_\theta$. $g_a$ and $g_s$ correspond to the main encoder and decoder, $h_a$ and $h_s$ to the hyper-encoder and hyper-decoder, and $C_m$ to the autoregressive context model described in \cite{minnen2018joint}. Skip connections represent element-wise additions between features. Q, AE and AD stand for quantization, arithmetic encoding, and arithmetic decoding steps, respectively. We fix $n=192$.}
\vspace{-0.1in}
\vspace{10pt}
\end{subfigure}
  \begin{subfigure}[t]{.49\textwidth}
  \centering
    \includegraphics[width=.9\textwidth]{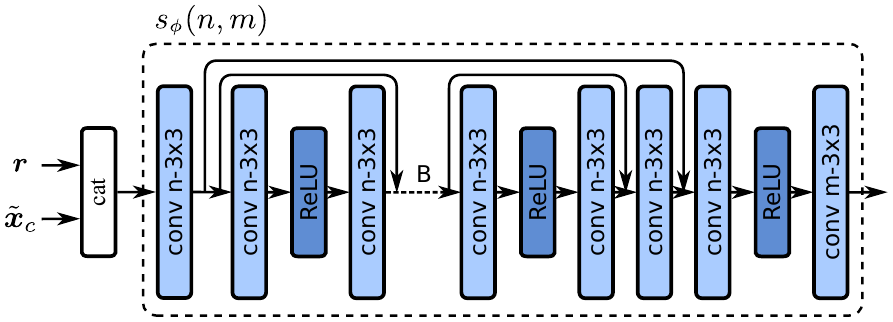}
\caption{Architecture details of the super-resolution network $s_\phi$. Skip connections stand for element-wise additions between features. }\label{fig:sr}
  \end{subfigure}\hfill
  \begin{subfigure}[t]{.49\textwidth}
  \centering
    \includegraphics[width=.9\textwidth]{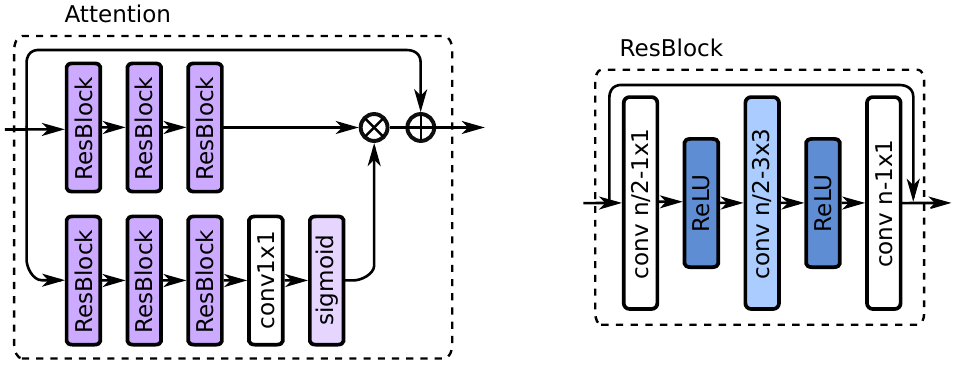}\caption{Building blocks of the conditional autoencoder $f_\theta$. These attention and residual blocks are implemented as proposed in \cite{cheng2020learned}.}\label{fig:other}
  \end{subfigure}
\caption{Architecture and details of CAESR.}
\label{fig:blocks}
\vspace{-0.2in}
\end{figure*}

\section{Proposed Solution}
\label{sec:proposed_solution}

\subsection{Framework and Formulation}

The overall pipeline of the proposed solution is described in Fig.~\ref{fig:proposed}. This hybrid layered system takes a low-resolution signal encoded with \gls{VVC} as a base-layer and a conditional \glsfirst{AE-HP}~\cite{balle2018variational} as an enhancement-layer that feeds a learning-based \gls{SR} module. In the following, let $s_\phi$ denotes the \gls{SR} module and $f_\theta$ the parametric function of the conditional autoencoder with hyperprior. 

Given an input image $\boldsymbol{x} \in \mathbb{R}^{W \times H \times 3}$ of width $W$ and height $H$, we first apply a spatial downscale by a factor 2 to generate the \gls{BL} images $\boldsymbol{x}_{lr}$. This latter is encoded with a \gls{VVC} encoder. The decoded  image $\boldsymbol{x}_c$ is then rescaled to the original resolution $W \times H$ to form the \gls{EL} model's input $\tilde{\boldsymbol{x}}_c$.
  
Our approach relies on conditional coding that allows a non-linear mixture of the source and the reconstructed \gls{BL} signal to be learned, improving the performance compared to residual coding \cite{ladune2021conditional}. Thus, the source image $\boldsymbol{x}$ and the upscaled base reconstruction $\tilde{\boldsymbol{x}}_c$ are concatenated along the feature axis to feed the autoencoder. The resulting tensor $(\tilde{\boldsymbol{x}}_c, \boldsymbol{x}) \in \mathbb{R}^{W \times H \times 6}$ is encoded by the encoder part of $f_\theta$, denoted as $g_a$, into a latent vector $\boldsymbol{y}$. Additional latent variables $\boldsymbol{z}$ are produced by the hyper-encoder $h_a$ to capture spatial dependencies among the element of $\boldsymbol{y}$. Both latents are quantized using the $round$ function to produce $\hat{\boldsymbol{y}}$ and $\hat{\boldsymbol{z}}$. At training, we apply a uniform noise $\mathcal{U}(-\frac{1}{2}, +\frac{1}{2})$ on latents to emulate the quantization errors while enabling backpropagation, resulting in $\tilde{\boldsymbol{y}}$ and $\tilde{\boldsymbol{z}}$. To simplify, we use $\bar{\boldsymbol{y}}$ and $\bar{\boldsymbol{z}}$ to denote both actual and emulated quantized latents. The latent variables are then entropy coded regarding a \gls{GMM} parameterized by the output of the hyper-decoder $h_s$ as:

\begin{equation}
    p(\bar{\boldsymbol{y}}|\bar{\boldsymbol{z}}) \sim \sum_{k=1}^{K} \boldsymbol{w}^{(k)}\mathcal{N}(\boldsymbol{\mu}^{(k)},\boldsymbol{\sigma}^{2(k)}),
    \label{eq:entropy}
\end{equation}

with $k$ the index of mixtures defined by $\boldsymbol{w}^{(k)}$, $\boldsymbol{\mu}^{(k)}$ and $\boldsymbol{\sigma}^{2(k)}$, denoting weights, means and scales, respectively.

At the decoder side, the latent residual signal $r$ is reconstructed by the synthesis part of $f_\theta$, denoted as $g_s$, and concatenated with the upscaled based-layer image $\tilde{\boldsymbol{x}}_c$ to form the input of the super-resolution network $s_\phi$. Finally, the output image $\hat{\boldsymbol{x}}_c$ is reconstructed from the following equation: 

\begin{equation}
    \hat{\boldsymbol{x}}_c = s_\phi(\tilde{\boldsymbol{x}}_c, r).
\end{equation}

In this work, the upscaling operation is applied before feeding the super-resolution module $s_\phi$ using an interpolation filter to make the network performing both high-resolution details recovering and conditional coding process inversion. 

All components of the overall differentiable system are jointly trained to minimize the following rate distortion loss function $\mathcal{L}$ based on a Lagrangian multiplier $\lambda$:

 \begin{equation}\label{eq:loss}
    \mathcal{L(\lambda)}=  D(\hat{\boldsymbol{x}}_c, \boldsymbol{x}) + \lambda R.
\end{equation}

The distortion $D$ is measured using the \gls{MSE} between $\hat{\boldsymbol{x}}_c$ and $\boldsymbol{x}$. The term $R$ corresponds to the Shannon entropy of $\tilde{\boldsymbol{y}}$, computed as:

\begin{equation}
    R = \mathbb{E}_{\tilde{\boldsymbol{y}} \sim m}[-\log_2(p(\tilde{\boldsymbol{y}}|\tilde{\boldsymbol{z}}))],
\end{equation}

with $m$ the true distribution of latents.

\subsection{Network Architecture}

The architecture of the proposed system, illustrated in Fig. \ref{fig:blocks}, is described in this section. 

\subsubsection{Autoencoder}

\begin{figure*}[t]
\centering
\includegraphics[width=\linewidth]{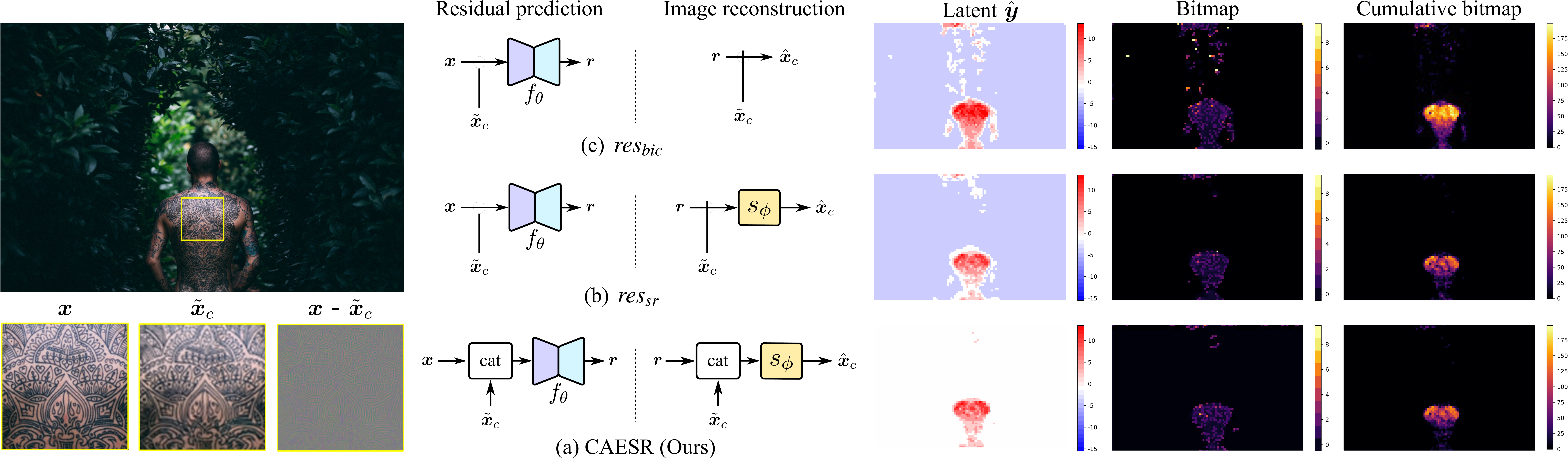}
\caption{Visualization of the configurations tested during ablation that include the autoencoder $f_\theta$ using \textit{felix-russell-saw-140699.png} from the CLIC21 validation set encoded with \acrshort{VTM} AI (qp27) as an example. We illustrate residual prediction and image reconstruction steps for each configuration. The latent $\hat{\boldsymbol{y}}$ and its corresponding bitmap represent the channel with the highest entropy. Cumulative bitmap corresponds to the sum of bitmaps over the latent channels.}
\label{fig:ablation_residual}
\vspace{-0.2in}
\end{figure*}

The structure of $f_\theta$ is based on the layered \gls{AE-HP} architecture described in \cite{cheng2020learned}, that estimates the group of parameters 
$\{\boldsymbol{w}^{(k)}, \boldsymbol{\mu}^{(k)},\boldsymbol{\sigma}^{2(k)}\}$, with $k=3$, for the entropy model described in (\ref{eq:entropy}). We also use an autoregressive context model over latents \cite{minnen2018joint}, denoted as $C_m$, to improve the entropy model accuracy without increasing the rate. The main analysis and synthesis transforms, $g_a$ and $g_s$, respectively, are composed of successive self-attention and residual blocks, depicted in Fig.~\ref{fig:other}. The non-linearity is integrated using the \gls{GDN} activation function \cite{balle2016end} and LeakyReLU as described in \cite{cheng2020learned}. For the hyper-encoder $h_a$ and hyper-decoder $h_s$, LeakyReLU activation function is used. Regarding dimensionality reduction and expansion strided convolutional layers and sub-pixel upscaling layers \cite{shi2016real} are implemented, respectively.

\subsubsection{Super-Resolution}
Our super-resolution module is inspired by the \gls*{EDSR} architecture~\cite{lim2017enhanced} which enables state-of-the-art performance. This \gls{SR} architecture mainly consists of $B$ \glspl*{RB} with short and long skip connections. In this work, we fix $B=8$ and use 64 filters of size $3 \times 3$ for each convolutional layer. We introduce the non-linearity with the ReLU activation, as described in Fig. \ref{fig:sr}. We removed the upscaling layer typically located at the end of the network and perform image upscaling before passing the input picture through this module.

\section{Experimental Results}
\label{sec:experimental_results}

\subsection{Training}
\label{subsec:training}

Both super-resolution and autoencoder networks are jointly trained to minimize the rate-distortion loss defined in (\ref{eq:loss}). 

We train our model using three different image datasets, namely DIV2K \cite{agustsson2017ntire}, Flickr2K, and the training dataset provided by the \gls{CLIC21} \cite{clic2021}. The performance is evaluated on the CLIC21 validation dataset, consisting of 42 images with various spatial characteristics. We first convert the image samples from PNG to YUV4:2:0 format. The base-layer input images $\boldsymbol{x}_{lr}$ are obtained by applying a bicubic downscale of factor 2. Then, the reconstructed versions $\tilde{\boldsymbol{x}}_c$ of the low-resolution images $\boldsymbol{x}_{lr}$, are obtained using the \gls*{VTM} in all-intra configuration for different \glspl{QP}. For simplicity, we generate YUV4:4:4 tensors by duplicating the chroma components for both reconstructed images $\tilde{\boldsymbol{x}}_c$ and original images $\boldsymbol{x}$, respectively. We crop $256 \times 256$ high-resolution and corresponding $128 \times 128$ low-resolution patches from the training set, resulting in around 150K training pairs. 

We train one model per base-layer $QP \in \{37,32,27,22\}$ and select specific $\lambda$ values in (\ref{eq:loss}). As the base quality is starting to saturate at higher bitrate, we empirically decided to allocate more bitrate for the lower \gls{BL} \glspl{QP}. The models are trained over a total of 20 epochs with a learning rate of $10^{-4}$. We apply a learning rate decay with a gamma of 0.5 for the last 5 epochs to improve the convergence. We use a batch size of 8 and optimize the model with ADAM~\cite{kingma2014adam} by setting $\beta_1=0.9$, $\beta_2=0.999$ and $\epsilon=10^{-8}$. For the whole experiments, the quality is assessed on the luma component using \gls*{PSNR} and \gls*{SSIM}~\cite{wang2004image} full-reference objective image quality metrics computed between the reconstructed images $\hat{\boldsymbol{x}}_c$ and original images $\boldsymbol{x}$.

\subsection{Ablation Study}

\begin{figure}[t]
\begin{subfigure}{0.49\linewidth}
\includegraphics[width=0.9\linewidth]{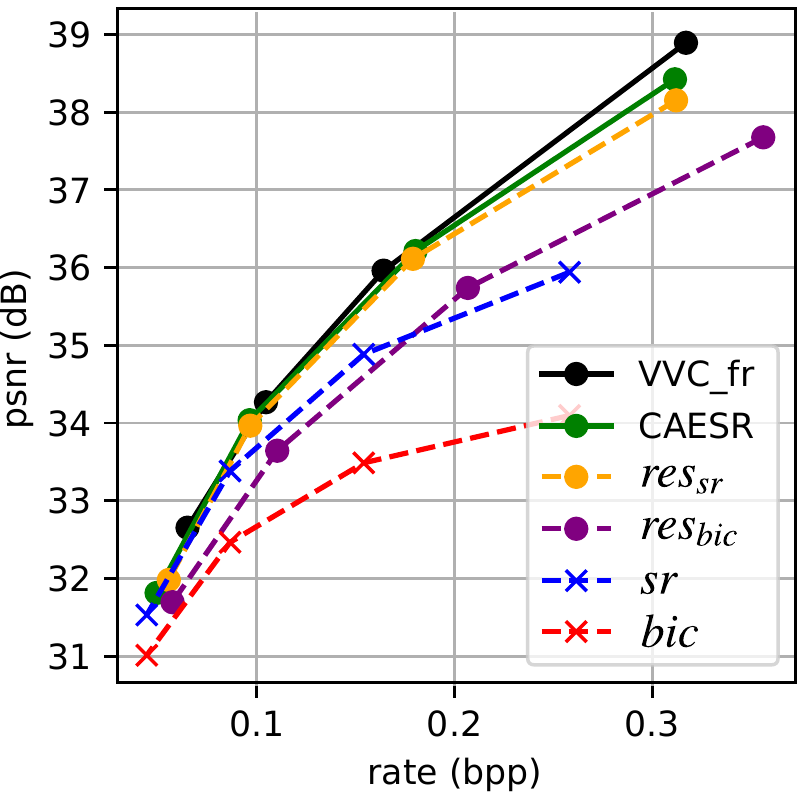}
\caption{PSNR.}\label{fig:psnr}
\end{subfigure}
\begin{subfigure}{0.49\linewidth}
\includegraphics[width=0.9\linewidth]{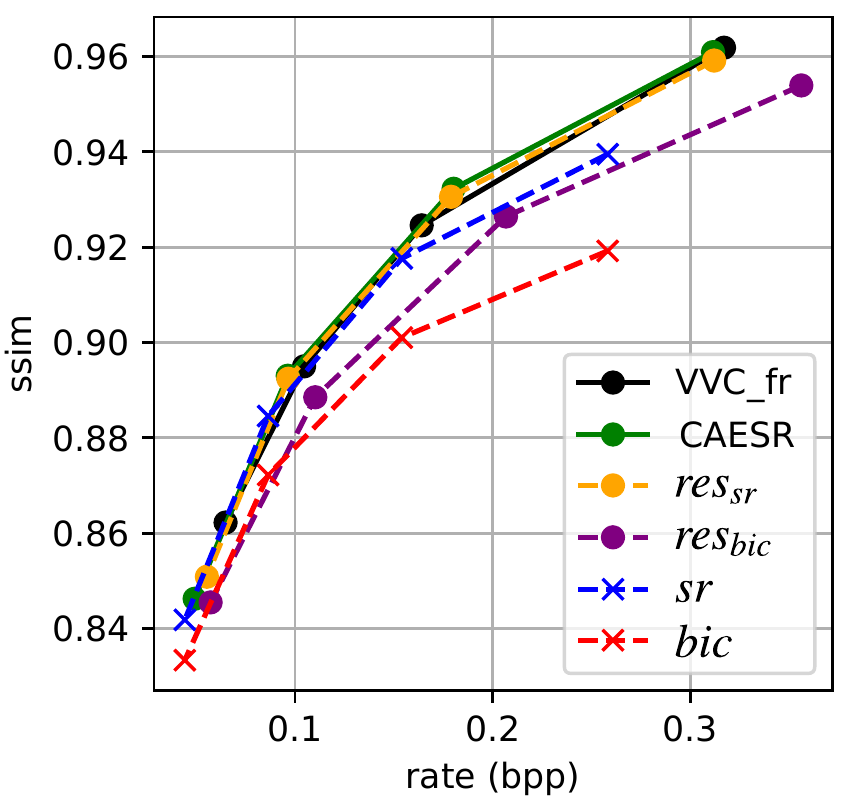}
\caption{SSIM.}\label{fig:ssim}
\end{subfigure}

\caption{Performance of the tested configurations on the CLIC2021 validation image dataset. We display incomplete systems for ablation study in dashed lines.}
\label{fig:ablation_rd}
\vspace{-0.2in}
\end{figure}

In this experiment, we demonstrate the effectiveness of the proposed system through an ablation study. The models that use our \gls{EL} module $f_{\theta}$, including the proposed conditional coding system CAESR and the residual-based configurations with and without super-resolution, represented by \textit{res\textsubscript{sr}} and \textit{res\textsubscript{bic}}, respectively, are illustrated in Fig. \ref{fig:ablation_residual}. For those configurations, we compute the global rate by summing both the \gls{BL} and \gls{EL} signal bitrates. This test also considers configurations based on our super-resolution module $s_\phi$ and a bicubic interpolation filter used as post-processing modules, represented by \textit{sr} and \textit{bic}, respectively. The whole learned models are optimized using the training strategy described in Section \ref{subsec:training}. 

We display latent variables and bitmaps obtained with the different tested models in the right part of Fig. \ref{fig:ablation_residual}. We observe that the configurations that include super-resolution, i.e., (a) and (b), produce more sparse latent variables that require fewer bits for enhancement layer encoding. The joint training of the super-resolution module $s_\phi$ and the autoencoder $f_\theta$ allows an optimal interaction between the two models. Therefore, high-frequencies that can be recovered by super-resolution are omitted by the autoencoder, allowing the autoencoder $f_\theta$ to focus on the most complex areas.

The \gls{RD} curves are represented in Fig. \ref{fig:ablation_rd}. We also add full-resolution single layer \gls{VVC} configuration, which corresponds to the high-resolution images encoded with \gls{VVC} \gls{VTM} all-intra mode. To match the bitrates obtained with our layered system, we select $QP \in \{42,39,36,31\}$ for full-resolution coding. The proposed conditional system outperforms all the other tested configurations in terms of rate-distortion performance, using the BD-BR (Bj{\o}ntegaard-Delta Bit-Rate) metric ~\cite{vceg_m33}, on the whole bitrate range. While offering spatial scalability, the BD-BR values of CAESR over the full-resolution coding anchor are 3.41\% and -3.49\% regarding the PSNR and SSIM metrics, respectively. We notice that the configurations that include both the autoencoder $f_\theta$ and the \gls{SR} module $s_\phi$ in the enhancement layer are more efficient than the others, particularly at higher bitrates. Indeed, in this range of bitrate, the reconstructed residual information contains high-resolution details that cannot be recovered using a single post-processing module. Although the residual bicubic configuration, i.e., (c) in Fig. \ref{fig:ablation_residual}, offers lower performance, this experiment demonstrates that at high bitrate, simply transmitting the residual with our system offers gains in PSNR over super-resolution used as a post-processing module.

\begin{figure}[t]
\centering
\includegraphics[width=\linewidth]{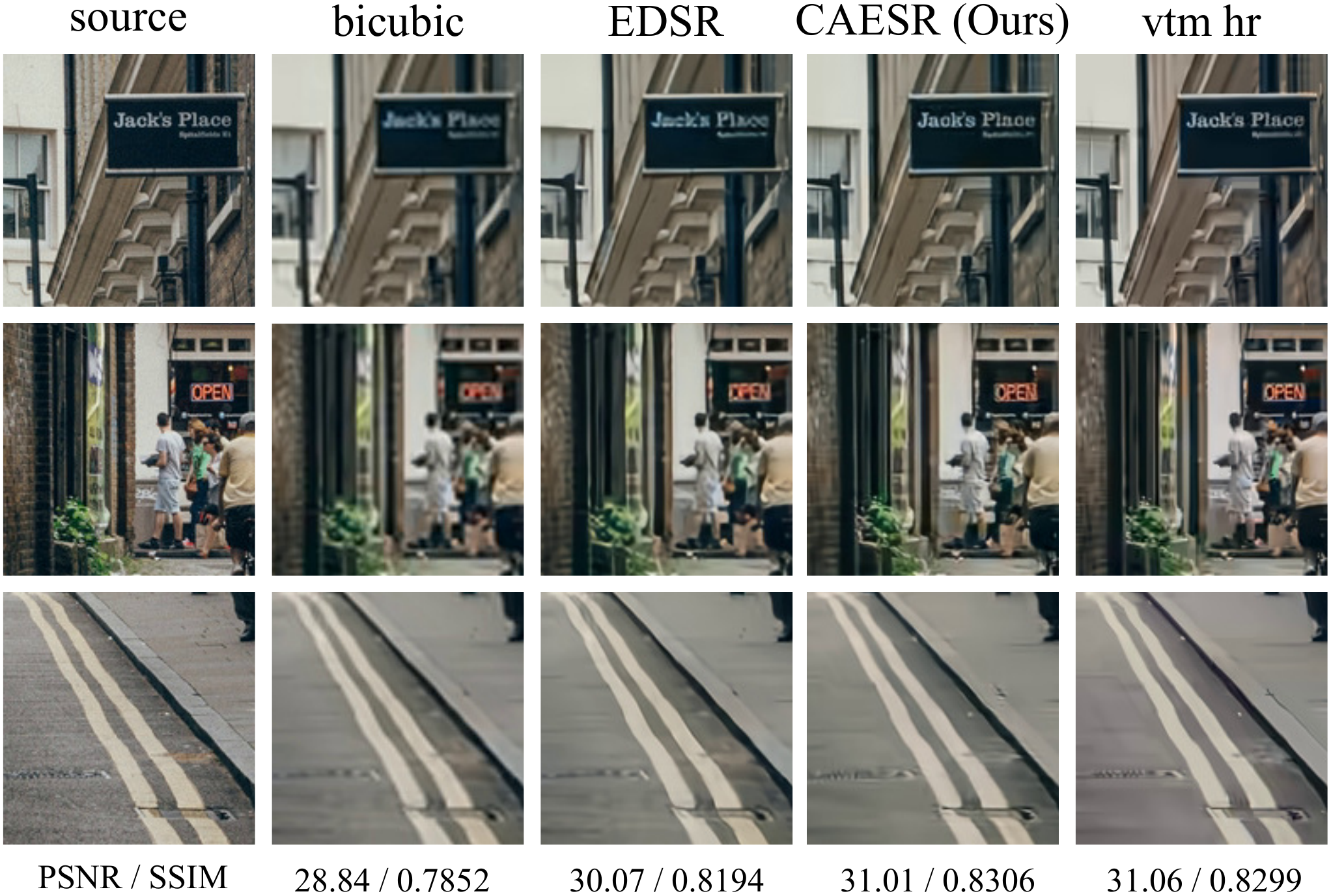}
\caption{Visual illustration of \textit{daniel-robert-405.png} (0.21bpp).}
\label{fig:daniel}
\vspace{-0.2in}
\end{figure}

\subsection{Visualization}

In this experiment, we visually compare our method against the super-resolution network \gls{EDSR} \cite{lim2017enhanced} used as a post-processing module on images. To allow a fair evaluation, we trained \gls{EDSR} following the experimental settings described in Section \ref{subsec:training}. For visualisation, we adjust the \gls{QP} of the \gls{EDSR} input to match the bitrate with our system. 

As depicted in Fig. \ref{fig:daniel}, our method produces better high-resolution images in terms of visual quality than both the bicubic filter and \gls{EDSR} used as post-processing, while being close to the full-resolution coding anchor. We observe that our system allows highly contrasted areas, like texts, to be accurately recovered from the low-resolution image.

\vspace{-0.05in}
\section{Conclusion}
\label{sec:conclusion}

In this paper, we present CAESR, an hybrid learning-based approach for spatial scalability based on the joint training of two deep \glspl{CNN}: a conditional autoencoder $f_\theta$ and a super-resolution module $s_\phi$. The deep autoencoder with hyperprior, learns to represent the residual information that cannot be recovered by the super-resolution module used as a post-processing step. This residual information is combined with the upscaled base-layer reconstruction at the decoder side to form the high-resolution output signal. Our approach relies on conditional coding that learns the optimal mixture of the source and the upscaled image, enabling better performance than residual coding. Our solution offers performances on par with \gls{VVC} full-resolution intra coding while being scalable.    

As future work, we plan to include the temporal aspect into our model to ensure inter-coded frame processing. 

\bibliographystyle{IEEEtran}
\bibliography{IEEEexample}

\end{document}

%% file: glossary.tex
\usepackage{hyperref}
\makeglossaries

\newacronym{VVC}{VVC}{versatile video coding}
\newacronym{HEVC}{HEVC}{high efficiency video coding}
\newacronym{QoE}{QoE}{quality of experience}
\newacronym{HDR}{HDR}{high dynamic range}
\newacronym{HFR}{HFR}{high frame-rate}
\newacronym{UHD}{UHD}{ultra high definition}
\newacronym{JVET}{JVET}{joint video exploration team}
\newacronym{ITU-T}{ITU-T}{international telecommunication union}
\newacronym{MPEG}{MPEG}{moving picture expert group}
\newacronym{AI}{AI}{artificial intelligence}
\newacronym{ANN}{ANN}{artificial neural network}
\newacronym{LR}{LR}{low-resolution}
\newacronym{HR}{HR}{high-resolution}
\newacronym{MTL}{MTL}{multitask learning}
\newacronym{QE}{QE}{quality enhancement}
\newacronym{SR}{SR}{super-resolution}
\newacronym{CR}{CR}{compact image-resolution}
\newacronym{QP}{QP}{quantization parameter}
\newacronym{VTM}{VTM-11}{VVC test model}
\newacronym{CTC}{CTC}{common test conditions}
\newacronym{BD}{BD}{Bjontegaard-Delta}
\newacronym{HD}{HD}{high-definition}
\newacronym{ML}{ML}{machine learning}
\newacronym{RB}{RB}{residual block}
\newacronym{EDSR}{EDSR}{enhanced deep super-resolution}
\newacronym{PSNR}{PSNR}{peak signal to noise ratio}
\newacronym{SSIM}{SSIM}{structural similarity}
\newacronym{BL}{BL}{base-layer}
\newacronym{EL}{EL}{enhancement-layer}
\newacronym{AE}{AE}{autoencoder}
\newacronym{HLS}{HLS}{high level syntax}
\newacronym{MSE}{MSE}{mean squared error}
\newacronym{GAN}{GAN}{generative adversarial networks}
\newacronym{SRA}{SRA}{spatial resolution adaptation}
\newacronym{CNN}{CNN}{convolutional neural network}
\newacronym{LCEVC}{LCEVC}{low complexity enhancement video coding}
\newacronym{GMM}{GMM}{gaussian mixture model}
\newacronym{GDN}{GDN}{generalized divisive normalization}
\newacronym{AE-HP}{AE-HP}{autoencoder with hyperprior}
\newacronym{CLIC21}{CLIC21}{challenge on learned image compression}
\newacronym{SHVC}{SHVC}{scalable high efficiency video coding}
\newacronym{RD}{RD}{rate-distortion}
\newacronym{bpp}{bpp}{bit per pixel}

%% file: conference_101719.bbl
\begin{thebibliography}{10}
\providecommand{\url}[1]{#1}
\csname url@samestyle\endcsname
\providecommand{\newblock}{\relax}
\providecommand{\bibinfo}[2]{#2}
\providecommand{\BIBentrySTDinterwordspacing}{\spaceskip=0pt\relax}
\providecommand{\BIBentryALTinterwordstretchfactor}{4}
\providecommand{\BIBentryALTinterwordspacing}{\spaceskip=\fontdimen2\font plus
\BIBentryALTinterwordstretchfactor\fontdimen3\font minus
  \fontdimen4\font\relax}
\providecommand{\BIBforeignlanguage}[2]{{%
\expandafter\ifx\csname l@#1\endcsname\relax
\typeout{** WARNING: IEEEtran.bst: No hyphenation pattern has been}%
\typeout{** loaded for the language `#1'. Using the pattern for}%
\typeout{** the default language instead.}%
\else
\language=\csname l@#1\endcsname
\fi
#2}}
\providecommand{\BIBdecl}{\relax}
\BIBdecl

\bibitem{boyce2015overview}
J.~M. Boyce, Y.~Ye, J.~Chen, and A.~K. Ramasubramonian, ``Overview of shvc:
  Scalable extensions of the high efficiency video coding standard,''
  \emph{IEEE Transactions on Circuits and Systems for Video Technology},
  vol.~26, no.~1, pp. 20--34, 2015.

\bibitem{maurer2020overview}
F.~Maurer, S.~Battista, L.~Ciccarelli, G.~Meardi, and S.~Ferrara, ``Overview of
  mpeg-5 part 2--low complexity enhancement video coding (lcevc),'' \emph{ITU
  Journal: ICT Discoveries}, vol.~3, no.~1, 2020.

\bibitem{bruckstein2003down}
A.~M. Bruckstein, M.~Elad, and R.~Kimmel, ``Down-scaling for better transform
  compression,'' \emph{IEEE Transactions on Image Processing}, vol.~12, no.~9,
  pp. 1132--1144, 2003.

\bibitem{zhang2019vistra2}
F.~Zhang, M.~Afonso, and D.~R. Bull, ``Vistra2: Video coding using spatial
  resolution and effective bit depth adaptation,'' \emph{Signal Processing:
  Image Communication}, p. 116355, 2021.

\bibitem{ma2019perceptually}
D.~Ma, M.~Afonso, F.~Zhang, and D.~R. Bull, ``Perceptually-inspired
  super-resolution of compressed videos,'' in \emph{Applications of Digital
  Image Processing XLII}, vol. 11137.\hskip 1em plus 0.5em minus 0.4em\relax
  International Society for Optics and Photonics, 2019, p. 1113717.

\bibitem{bonnineau2020versatile}
C.~Bonnineau, W.~Hamidouche, J.-F. Travers, and O.~Deforges, ``Versatile video
  coding and super-resolution for efficient delivery of 8k video with 4k
  backward-compatibility,'' in \emph{ICASSP 2020-2020 IEEE International
  Conference on Acoustics, Speech and Signal Processing (ICASSP)}.\hskip 1em
  plus 0.5em minus 0.4em\relax IEEE, 2020, pp. 2048--2052.

\bibitem{bonnineau2021}
C.~Bonnineau, W.~Hamidouche, J.-F. Travers, N.~Sidaty, and O.~Deforges,
  ``Multitask learning for vvc quality enhancement and super-resolution,'' in
  \emph{2021 Picture Coding Symposium (PCS)}, 2021, pp. 1--5.

\bibitem{toderici2015variable}
\BIBentryALTinterwordspacing
G.~Toderici, S.~M. O'Malley, S.~J. Hwang, D.~Vincent, D.~Minnen, S.~Baluja,
  M.~Covell, and R.~Sukthankar, ``Variable rate image compression with
  recurrent neural networks,'' in \emph{International Conference on Learning
  Representations}, 2016. [Online]. Available:
  \url{http://arxiv.org/abs/1511.06085}
\BIBentrySTDinterwordspacing

\bibitem{balle2016end}
``\BIBforeignlanguage{English (US)}{End-to-end optimized image compression},''
  2017, 5th International Conference on Learning Representations, ICLR 2017 ;
  Conference date: 24-04-2017 Through 26-04-2017.

\bibitem{balle2018variational}
\BIBentryALTinterwordspacing
J.~Ballé, D.~Minnen, S.~Singh, S.~J. Hwang, and N.~Johnston, ``Variational
  image compression with a scale hyperprior,'' in \emph{International
  Conference on Learning Representations}, 2018. [Online]. Available:
  \url{https://openreview.net/forum?id=rkcQFMZRb}
\BIBentrySTDinterwordspacing

\bibitem{cheng2020learned}
Z.~Cheng, H.~Sun, M.~Takeuchi, and J.~Katto, ``Learned image compression with
  discretized gaussian mixture likelihoods and attention modules,'' in
  \emph{Proceedings of the IEEE/CVF Conference on Computer Vision and Pattern
  Recognition}, 2020, pp. 7939--7948.

\bibitem{lu2019dvc}
G.~Lu, W.~Ouyang, D.~Xu, X.~Zhang, C.~Cai, and Z.~Gao, ``Dvc: An end-to-end
  deep video compression framework,'' in \emph{Proceedings of the IEEE/CVF
  Conference on Computer Vision and Pattern Recognition}, 2019, pp.
  11\,006--11\,015.

\bibitem{tsai2018learning}
Y.-H. Tsai, M.-Y. Liu, D.~Sun, M.-H. Yang, and J.~Kautz, ``Learning binary
  residual representations for domain-specific video streaming,'' in
  \emph{Proceedings of the AAAI Conference on Artificial Intelligence},
  vol.~32, no.~1, 2018.

\bibitem{akbari2019dsslic}
M.~Akbari, J.~Liang, and J.~Han, ``Dsslic: deep semantic segmentation-based
  layered image compression,'' in \emph{ICASSP 2019-2019 IEEE International
  Conference on Acoustics, Speech and Signal Processing (ICASSP)}.\hskip 1em
  plus 0.5em minus 0.4em\relax IEEE, 2019, pp. 2042--2046.

\bibitem{lee2020hybrid}
W.-C. Lee, C.-P. Chang, W.-H. Peng, and H.-M. Hang, ``A hybrid layered image
  compressor with deep-learning technique,'' in \emph{2020 IEEE 22nd
  International Workshop on Multimedia Signal Processing (MMSP)}.\hskip 1em
  plus 0.5em minus 0.4em\relax IEEE, 2020, pp. 1--6.

\bibitem{minnen2018joint}
D.~Minnen, J.~Ball{\'e}, and G.~D. Toderici, ``Joint autoregressive and
  hierarchical priors for learned image compression,'' \emph{Advances in Neural
  Information Processing Systems}, vol.~31, pp. 10\,771--10\,780, 2018.

\bibitem{ladune2021conditional}
T.~Ladune, P.~Philippe, W.~Hamidouche, L.~Zhang, and O.~D{\'e}forges,
  ``Conditional coding for flexible learned video compression,'' in
  \emph{International Conference on Learning Representations (ICLR) 2021,
  Neural Compression Workshop}, 2021.

\bibitem{shi2016real}
W.~Shi, J.~Caballero, F.~Husz{\'a}r, J.~Totz, A.~P. Aitken, R.~Bishop,
  D.~Rueckert, and Z.~Wang, ``Real-time single image and video super-resolution
  using an efficient sub-pixel convolutional neural network,'' in
  \emph{Proceedings of the IEEE conference on computer vision and pattern
  recognition}, 2016, pp. 1874--1883.

\bibitem{lim2017enhanced}
B.~Lim, S.~Son, H.~Kim, S.~Nah, and K.~Mu~Lee, ``Enhanced deep residual
  networks for single image super-resolution,'' in \emph{Proceedings of the
  IEEE conference on computer vision and pattern recognition workshops}, 2017,
  pp. 136--144.

\bibitem{agustsson2017ntire}
E.~Agustsson and R.~Timofte, ``Ntire 2017 challenge on single image
  super-resolution: Dataset and study,'' in \emph{Proceedings of the IEEE
  Conference on Computer Vision and Pattern Recognition Workshops}, 2017, pp.
  126--135.

\bibitem{clic2021}
C.~on~Learned Image~Compression, ``https://www.compression.cc/,'' June 2021.

\bibitem{kingma2014adam}
\BIBentryALTinterwordspacing
D.~P. Kingma and J.~Ba, ``Adam: {A} method for stochastic optimization,'' in
  \emph{3rd International Conference on Learning Representations, {ICLR} 2015,
  San Diego, CA, USA, May 7-9, 2015, Conference Track Proceedings}, Y.~Bengio
  and Y.~LeCun, Eds., 2015. [Online]. Available:
  \url{http://arxiv.org/abs/1412.6980}
\BIBentrySTDinterwordspacing

\bibitem{wang2004image}
Z.~Wang, A.~C. Bovik, H.~R. Sheikh, and E.~P. Simoncelli, ``Image quality
  assessment: from error visibility to structural similarity,'' \emph{IEEE
  transactions on image processing}, vol.~13, no.~4, pp. 600--612, 2004.

\bibitem{vceg_m33}
G.~Bj{\o}ntegaard, ``Document {VCEG-M33 ITU-T Q6/16}: {Calculation} of
  {Average} {PSNR} {Differences} {Between} {RD- Curves},'' April 2001.

\end{thebibliography}
